\documentclass[aps,reprint,superscriptaddress,secnumarabic,amssymb,showpacs]{revtex4-1}

\usepackage{bm}% bold math
\usepackage{dcolumn}
\usepackage{color}
\usepackage{amsmath}
\usepackage{adjustbox}
\usepackage{multirow}
\usepackage{graphicx}
\begin{document}

\title{Transistor concepts based on lateral heterostructures \\ of metallic and semiconducting phases of MoS$_2$}

\author{Damiano Marian}
\affiliation{Dipartimento di Ingegneria dell'Informazione, Universit\`a di Pisa\\
Via G. Caruso 16, 56122, Pisa, Italy}
\author{Elias Dib}
\affiliation{Dipartimento di Ingegneria dell'Informazione, Universit\`a di Pisa\\
Via G. Caruso 16, 56122, Pisa, Italy}
\author{Teresa Cusati}
\affiliation{Dipartimento di Ingegneria dell'Informazione, Universit\`a di Pisa\\
Via G. Caruso 16, 56122, Pisa, Italy}
\author{Enrique G. Marin}
\affiliation{Dipartimento di Ingegneria dell'Informazione, Universit\`a di Pisa\\
Via G. Caruso 16, 56122, Pisa, Italy}
\author{Alessandro Fortunelli}
\affiliation{CNR-ICCOM, Istituto di Chimica dei Composti Organometallici \\
Via G. Moruzzi 1, 56124, Pisa, Italy}
\author{Giuseppe Iannaccone}
\email{g.iannaccone@unipi.it}
\affiliation{Dipartimento di Ingegneria dell'Informazione, Universit\`a di Pisa\\
Via G. Caruso 16, 56122, Pisa, Italy}
\author{Gianluca Fiori}
\affiliation{Dipartimento di Ingegneria dell'Informazione, Universit\`a di Pisa\\
Via G. Caruso 16, 56122, Pisa, Italy}

\begin{abstract}
In this paper we propose two transistor concepts based on lateral heterostructures of monolayer MoS$_2$, composed of adjacent regions of 1T (metallic) and 2H (semiconducting) phases, inspired by recent research showing the possibility to obtain such heterostructures by electron beam irradiation.
The first concept, the lateral heterostructure field-effect transistor, exhibits potential of better performance with respect to the foreseen evolution of CMOS technology, both for high performance and low power applications. Performance potential has been evaluated by means of detailed multi-scale materials and device simulations. The second concept, the planar barristor, also exhibits potential competitive performance with CMOS, and an improvement of orders of magnitude in terms of the main figures of merit with respect to the recently proposed vertical barristor.
\end{abstract}

\maketitle

\section{Introduction}
Engineering new materials with tailored properties for electronic applications is the quintessential frontier of electronics.
It was embodied in the ``bandgap engineering" \cite{Capasso1987} or ``band structure engineering" \cite{Yablonovitch1987} paradigm of III-V materials systems in the eighties, and is embodied now in the ``materials on demand" paradigm \cite{novoselov2012} of ``van der Waals heterostructures" \cite{Geim2013}, where 2D materials of incommensurable lattice, including a broad set of atomic species, are stacked to create new 3D materials.

One can also have ``lateral heterostructures" \cite{Ci2010}, as in the example by Levendorf et al. \cite{Levendorf2012}, where a planar heterostructure of graphene and hexagonal boron nitride has been fabricated by successive graphene growth, patterning, and boron nitride regrowth in the same 2D sheet, exploiting the almost ideal lattice match. 

With lateral and vertical heterostructures of 2D materials the prospect emerges of combining materials with different electronic properties, such as semiconductors, insulators, and metals, in order to 
engineer materials for particular applications.

In this paper we focus on monolayer MoS$_2$. The advantage of designing devices made of 2D materials relies on an excellent electrostatic control of the channel by the gate. MoS$_2$ has also a direct bandgap between 1.5 and 1.8 eV, which makes it favorable for digital applications against other 2D materials, such as graphene that does not present a gap. MoS$_2$ has already shown remarkable electronic and optoelectronic properties \cite{Rad2011}, including a higher mobility than ultrathin silicon and germanium, and the possibility to be used in flexible electronic circuits, that is currently out of reach for silicon technology \cite{Wang2012}.

Molybdenum disulphide has also been used in vertical heterostructure devices, such as atomically thin p-n heterojunctions \cite{Lee2014} combining monolayer n-type MoS$_2$ and p-type WSe2, which shows a rectifying behavior. Also lateral heterostructures based on junctions of 1T and 2H phases have been proposed experimentally showing interesting and promising properties \cite{Eda2012,Kappera2014,Kappera2014_2}.

An intriguing property of MoS$_2$, which is decisive for the realization of lateral heterostructures, is that by high-dose electron beam irradiation one can induce a phase transition from the semiconducting phase (2H) to the metallic phase (1T), in a size-controllable region \cite{Lin2014}. This enables, in principle, top-down patterning of MoS$_2$ lateral 2H-1T heterostructures, whose in-plane charge transport properties critically depend on the quality of the lateral heterointerface.

The first demonstration of a Schottky diode, through the patterning of a lateral heterostructure with a metal region (1T-MoS$_2$) and a semiconducting region (2H-MoS$_2$), has already been reported \cite{Cusati2016}. 
Here, we exploit the possibility of fabricating lateral heterostructures in single-layer MoS$_2$ to explore two device concepts, evaluating by multiscale simulations its potential performances.

\begin{figure} [h!!!]
\includegraphics[width=8.5cm]{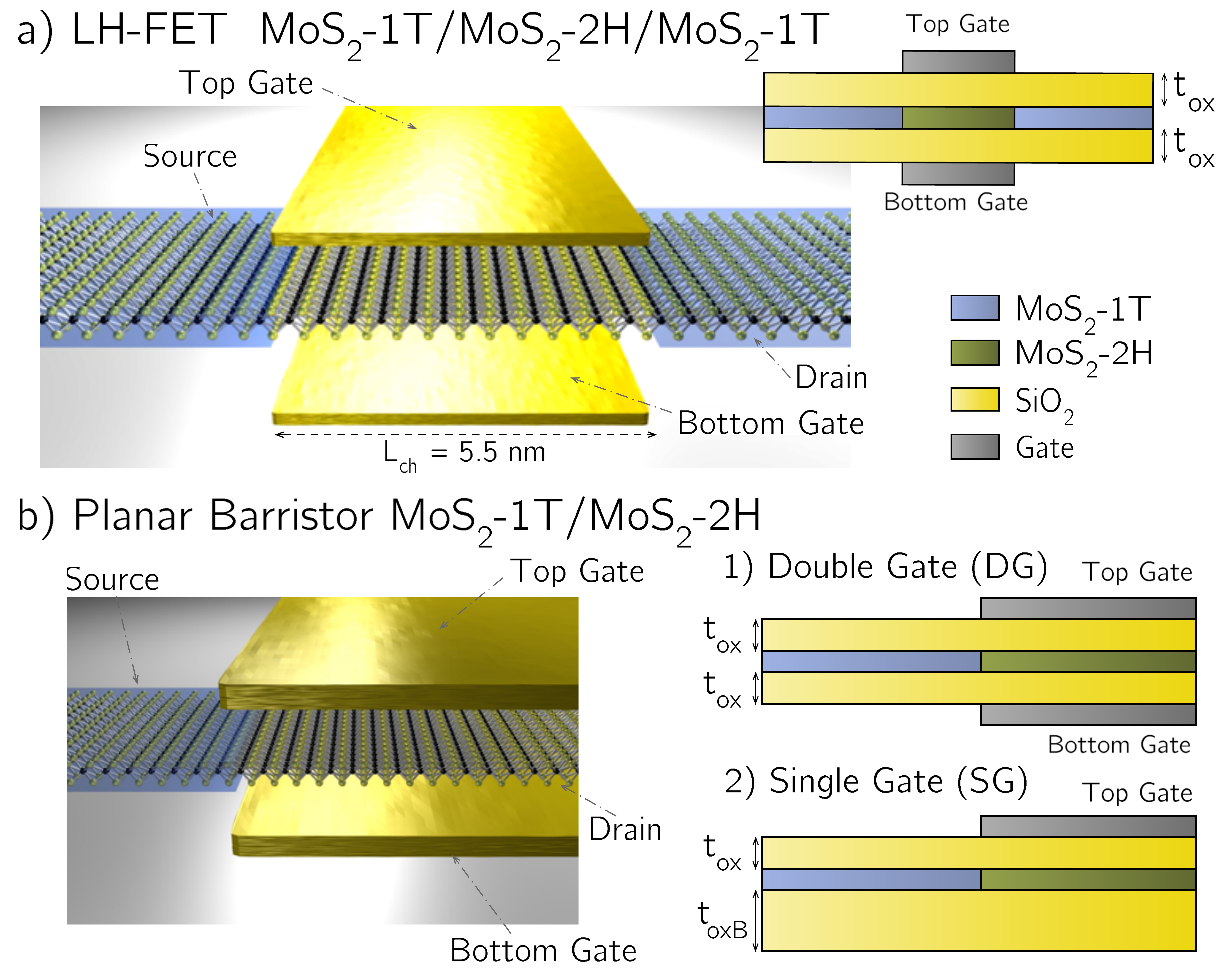}
\caption{a) MoS$_2$-2H/MoS$_2$-1T/MoS$_2$-2H double gate FET. Source, drain, bottom gate, top gate and channel length ($L_{\text{ch}}$~=~5.5 nm) are indicated. The oxide is SiO$_2$, with thickness $t_{\text{ox}}$~=~0.5 nm. In the inset, a longitudinal cross-section of the device is shown. b) Double gate MoS2-1T/MoS2-2H planar barristor. In the inset b1), a longitudinal cross-section of the double-gate device is shown, with $t_{\text{ox}}~=~0.5$~nm. In the inset b2) a longitudinal cross section of a single gate configuration, where $t_\text{oxB}$ is the bottom oxide thickness.}
\label{fig:Fig2}
\end{figure}

The first transistor concept is a lateral heterostructure field-effect transistor (LH-FET), shown in Fig. \ref{fig:Fig2}(a): It is a double-gate FET in which the channel consists of single-layer molybdenum disulfide. We have the semiconducting phase in the central region aligned with the gate, and the metallic phase in the source and drain extensions. The gate dielectric is silicon oxide with a thickness of 0.5~nm. It can be seen as an embodiment of the LH-FET proposed in 2011 for the graphene-boron-nitride heterostructure \cite{Fiori2011,Fiori2012}.

The second transistor concept is a lateral gated Schottky diode (Fig.~\ref{fig:Fig2}(b)), with a metallic source (1T-MoS$_2$) and a semiconducting drain (2H-MoS$_2$). We consider a double gate structure with a gate oxide thickness $t_{\text{ox}}$ (Fig.~\ref{fig:Fig2}(b1)) and a single (top) gate configuration with a top oxide thickness $t_{\text{ox}}$ and a bottom oxide thickness $t_\text{oxB}$ (Fig.~\ref{fig:Fig2}(b2)). The gate voltage modulates the height of the Schottky barrier and therefore the current. We call this transistor ``planar barristor", because the operating principle is similar to that of the vertical barristor proposed in 2012 \cite{Yang2012}.

\section{Theory}

In order to explore the viability and potential performance of such devices, we adopt a first-principle multi-scale modeling approach \cite{Bruzzone2014}, articulated in three steps. We start from Density Functional Theory (DFT) simulations of the electronic properties of 1T-2H heterostructures, from which we obtain a Hamiltonian defined on a basis of plane waves. From this we then extract a Hamiltonian on a basis of Maximally Localized Wannier Functions (MLWF) basis set. Finally, the resulting Hamiltonian is used in the open source device simulator NanoTCAD ViDES \cite{ViDES}, based on a Non-Equilibrium Green's Functions formalism, to investigate the device performance for digital applications. 

{\em Ab initio} calculations have been performed with Quantum Espresso \cite{QE}, using a 8x12x1 Monkhorst-Pack grid, ultrasoft pseudopotentials and the PBE exchange correlation functional. A 60 Ry wave function cutoff, a 600 Ry charge density cutoff and a vacuum layer of 20 \r{A}, to avoid interactions with periodic replicas, have been considered. The band structure and the Hamiltonian in the MLWF basis have been computed with Wannier90 \cite{Wannier} using the same grid as in DFT calculations. The obtained tight-binding-like Hamiltonian has then been included in the open source device simulator NanoTCAD ViDES \cite{ViDES}. In particular, we have simplify the full 22-band electronic structure model of Fig.~\ref{fig:Fig1} into a 6-band model which provides the same results for energies close the Fermi energy, but with reduced computational requirements (see Appendix). Then we have included the 6-band tight-binding-like Hamiltonian into the ViDES framework to compute transport in far-from-equilibrium conditions. In order to connect the two different phases of MoS$_2$ along the transport direction, the off diagonal elements of the Hamiltonian has chosen to be equal to MoS$_2$-2H elements (details of the method are reported in the Appendix). For all devices we have considered operation at temperature of 300~K and fully coherent quantum transport.

The complete multi-scale approach is thoroughly discussed in Bruzzone {\it et al.} \cite{Bruzzone2014} and in the Appendix.

\section{Results and Discussion}

\begin{figure} [h!!!]
\includegraphics[width=8.5cm]{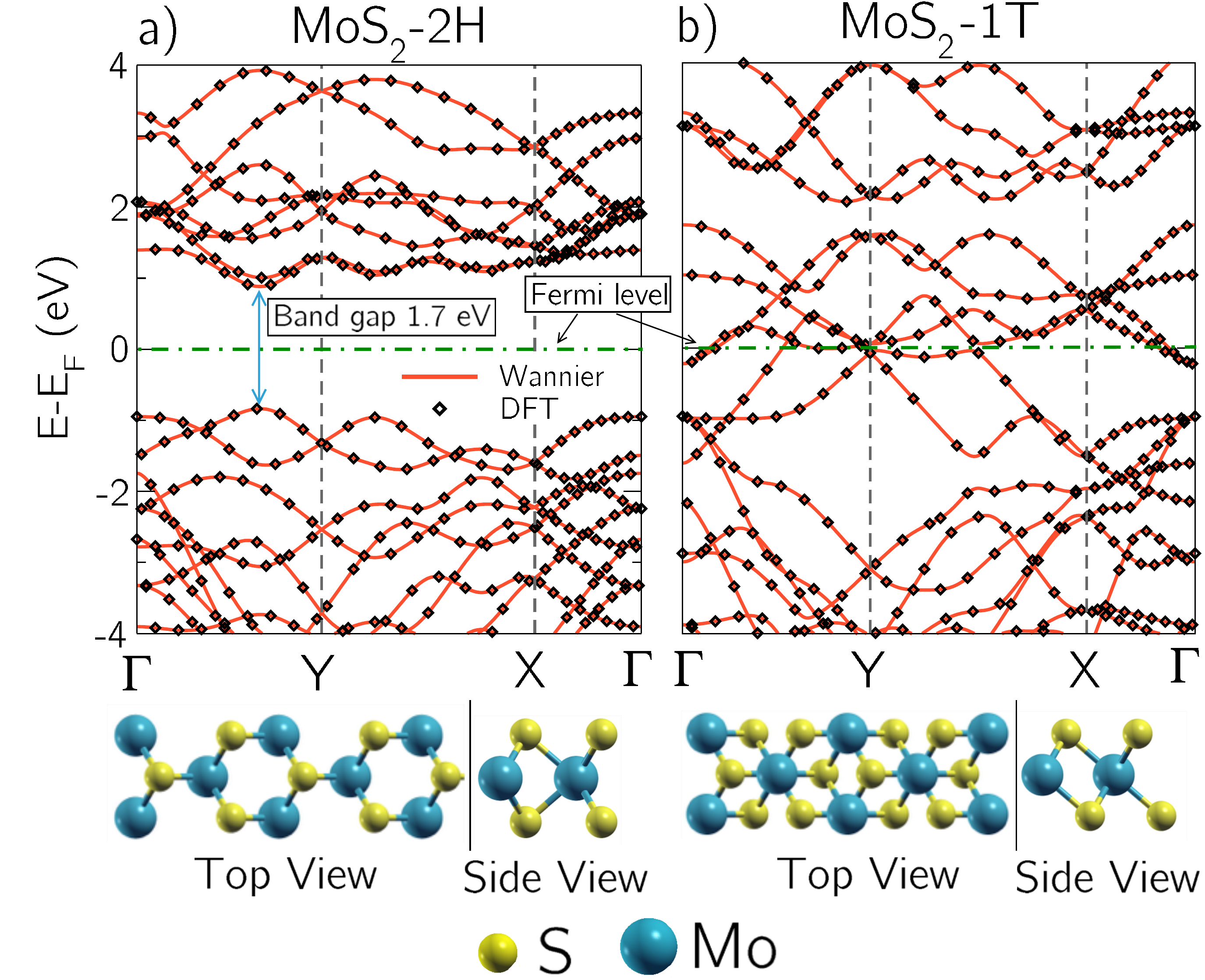}
\caption{Bands of a) MoS$_2$-2H and b) MoS$_2$-1T b) computed with DFT calculations (black dots) and with MLWF (red solid lines, 22 bands). Below each figure the schematic top and side views of the two different atomic-structured monolayer of MoS$_2$-2H a) and MoS$_2$-1T b) are shown.}
\label{fig:Fig1}
\end{figure}

The band structures of MoS$_2$-2H and MoS$_2$-1T are shown in Fig.~\ref{fig:Fig1}a) and \ref{fig:Fig1}b), respectively, where dots correspond to DFT calculations, and dashed lines to results using the MLWF basis. As apparent  from Fig.~\ref{fig:Fig1}, MoS$_2$-2H has a gap of 1.7 eV, whereas the 1T phase shows metallic behavior. 

\begin{figure} [h!!!]
\includegraphics[width=9cm]{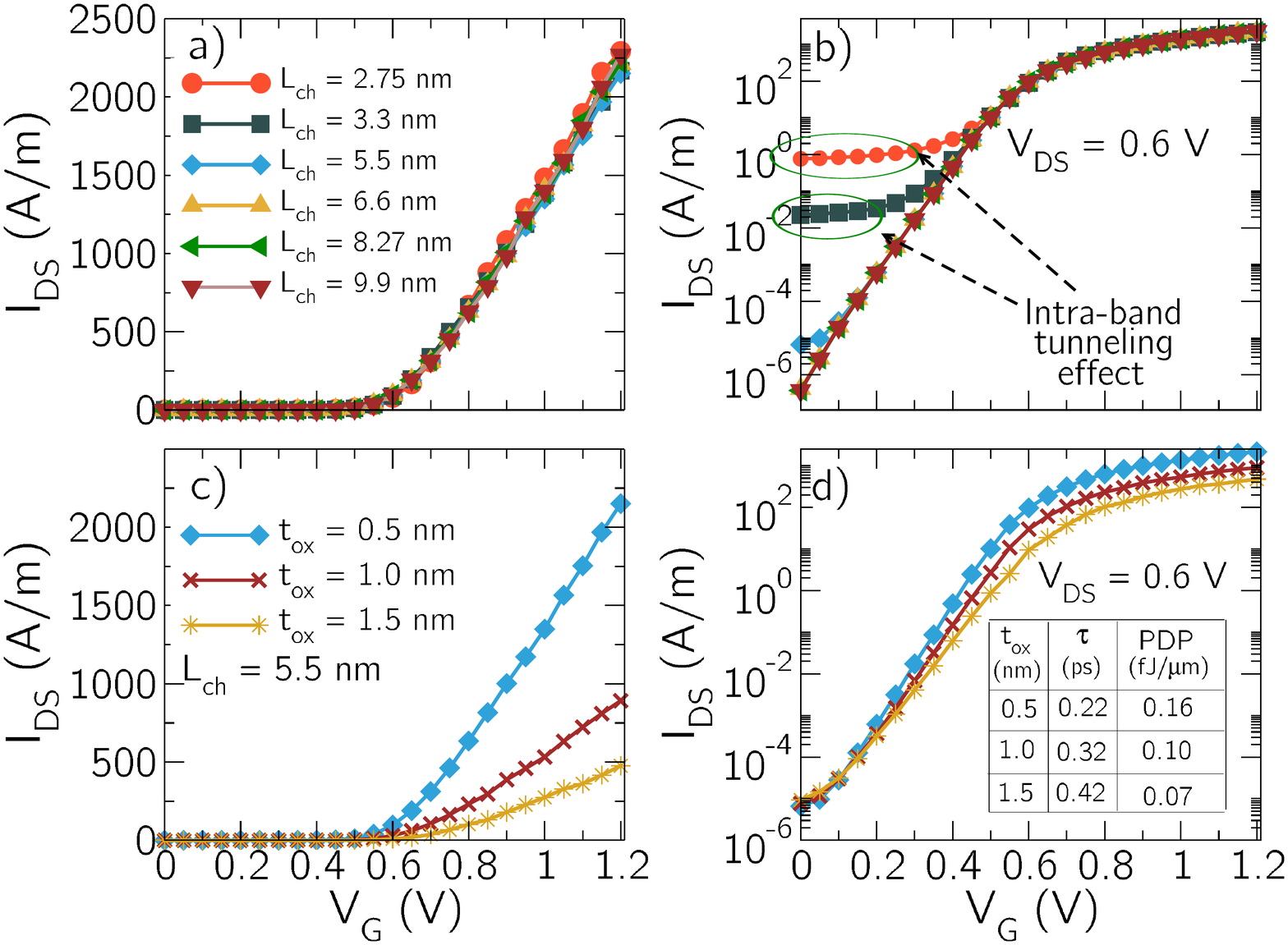}
\caption{Transfer characteristics in a) linear and b) semi-logarithmic scale of the device in Fig. \ref{fig:Fig2}a), for different MoS$_2$-2H channel lengths and for $V_{\text{DS}}$~=~0.6~V for an oxide thickness of $\text{t}_{\text{ox}}$ = 0.5 nm; and c) linear and d) semi-logarithmic for a MoS$_2$-2H channel length of 5.5 nm, $V_{\text{DS}}$~=~0.6~V and for different oxide thicknesses. In the inset $\tau$ and PDP, calculated for high performance process, are reported.}
\label{fig:Fig3}
\end{figure}

The transfer characteristics (I$_{\text{DS}}$-$V_{\text{G}}$) of the LH-FET are shown in Figs. \ref{fig:Fig3}a) in linear scale and \ref{fig:Fig3}b) in semilogarithmic scale for MoS$_2$-2H channel/gate lengths ranging from 2.75 nm to 9.9 nm. The applied drain-to-source voltage  $V_{\text{DS}}$ is 0.6~V and the oxide thickness $t_{\text{ox}}$ is 0.5~nm. In the case of the shorter channel lengths (2.75~nm and 3.3~nm) intra-band tunneling dominates the current and imposes a lower limit on the off-state current.

\begin{table*}
  \caption{\ Figures of merit for different channel lengths of the LH-FET and DG Planar Barristor for HP and LP}
  \label{tab1}
   \begin{tabular*}{\textwidth}{ l l | l  l l l l | l  l l l l }
   \hline
\multirow{3}{*}{}  &  &  \multicolumn{5}{l}{High Performance}   &  \multicolumn{4}{|l}{Low Power} \\
\hline
 &    &  SS & I$_{\text{on}}$/I$_{\text{off}}$  & $\tau$ & PDP & f$_T$    & SS & I$_{\text{on}}$/I$_{\text{off}}$  & $\tau$ & PDP \\
  &  & (mV/dec) &  & (ps) & (fJ/$\mu$m) & (THz)   & (mV/dec) &  & (ps) & (fJ/$\mu$m)  \\
\hline
\multirow{3}{*}{LH-FET}  & L$_{\text{ch}}$ = 3.3 nm     & 102  & $1.0 \times 10^4$  & 0.16  & 0.10 & 3  & * & *  & *  & * \\
\multirow{3}{*}{(V$_{\text{DS}}$ = 0.6 V)}  &  L$_{\text{ch}}$ = 5.5 nm     & 69  & $1.18 \times 10^4$  & 0.22  & 0.16 & 2  \ & 72 & $4.38 \times 10^6$ &  0.44 & 0.12 \\
  & L$_{\text{ch}}$ = 6.6 nm      & 69  & $1.20 \times 10^4$  & 0.25  & 0.18 & 1.5  &70 & $4.43 \times 10^6$ & 0.49 & 0.13 \\
  & L$_{\text{ch}}$ = 8.27 nm     & 69  & $1.23 \times 10^4$  & 0.30  & 0.22 & 1.4  & 68 & $4.65 \times 10^6$ & 0.47 & 0.13 \\
 & L$_{\text{ch}}$ = 9.9 nm     & 69  & $1.25 \times 10^4$  & 0.35  & 0.26 & 1.0  & 69 & $4.41 \times 10^6$ & 0.61 & 0.16  \\
\hline
Planar  & DG     & 68.5  & $9.8 \times 10^3$  & 0.14  & 0.055 & 3   & 72.5 & $3.5 \times 10^5$  & 1.2 & 0.017  \\
Barristor  & SG (t$_{\text{oxB}}$ \!\! = \!\! 0.5 nm)    & 73  & $4.4 \times 10^3$  & 0.16  & 0.028 & 2.8  & 79 & $1.5 \times 10^5$ & 1.8 &  0.011   \\
(V$_{\text{DS}}$ = 0.4 V) \ & SG (t$_{\text{oxB}}$ \!\! = \!\! 5 nm)     & 79  & $3.7 \times 10^3$  & 0.2  & 0.029 & 2.3  & 85 & $7.4 \times 10^4$ & 4.4 & 0.013  \\
\hline
  \end{tabular*}
  \begin{flushleft}
  *For channel length of 3.3~nm I$_{\text{off}}$ is not defined for LP case because of the high degradation of the OFF current due to the intra-band tunneling effect (see main text for further information).
\end{flushleft}
\end{table*}

To assess device performance, we use as a benchmark the consensus on the evolution of CMOS technology represented by the 
2015 edition of the International Technology Roadmap for Semiconductors (ITRS) \cite{ITRS}. The ITRS considers that CMOS processes are optimized either for High Performance (HP) or for Low Power (LP) applications, and therefore have different threshold voltages. 
To enable comparison with the ITRS, we consider two main bias conditions for the nMOSFET: the OFF state, corresponding to $V_{\rm GS}~=~V_{\rm off}$ and $V_{\rm DS} = V_{\rm DD}$, where $V_{\text{DD}}$ is the supply voltage; and the ON state, corresponding to $V_{\rm GS} = V_{\rm off} + V_{\rm DD}$ and $V_{\rm DS} = V_{\rm DD}$. In the case of HP process $V_{\rm off}$ is defined as the gate voltage for which the drain current in the OFF state is $I_{\rm off} = 100$~nA/$\mu$m. In case of LP process, in order to reduce the static power consumption, $V_{\rm off}$ is defined as the gate voltage for which the device current is $I_{\rm off} = 100$~pA/$\mu$m. Typically, the transfer characteristics is shifted by properly tuning the gate workfunction for each process in order to obtain $V_{\rm off} = 0$~V.

For all the channel lengths of the LH-FETs we consider the same supply voltage $V_{\rm DD} = 0.6$~V and the same oxide thickness $t_{\rm ox}$. The main figures of merit for the two process optimizations are reported in Table \ref{tab1}. The subthreshold swing (SS) is defined as the inverse slope of the I$_{\text{DS}}$-V$_{\text{GS}}$ curve in semi-logarithmic scale in the subthreshold regime. We also report the $I_{\text{on}}$/$I_{\text{off}}$ ratio, where $I_{\rm on}$ is the current in the ON state. As aforementioned the 2.75~nm and 3.3~nm FETs have a large intra-band tunneling current in the OFF state, so that $I_{\text{off}}$ cannot be defined for the LP case for L$_{\text{ch}}$~=~3.3~nm while for L$_{\text{ch}}$~=~2.75~nm $I_{\text{off}}$ cannot be defined for both LP and HP cases. We remark that, the bandgap calculated through DFT is underestimated, with a GW correction the band gap could increase to 2.1 eV, slightly reducing the intra-band tunneling effects observed in the shortest channels.

In Table \ref{tab1} we also show the intrinsic delay time $\tau$, which is a measure of the switching speed (a rough estimation of the switching time of a CMOS inverter), defined as:

\begin{equation}
\tau = \frac{Q_{\text{on}}-Q_{\text{off}}}{I_{\text{on}}} 
\end{equation}

where $Q_{\text{on}}$ and $Q_{\text{off}}$ are the total charge in the channel in the ON and OFF states, respectively; as well as the power delay product (PDP), which is a measure of energy efficiency (roughly proportional to the energy required to switch a logic gate),

\begin{equation}
PDP = V_{\text{DD}}I_{\text{on}}\tau = V_{\text{DD}}\left(Q_{\text{on}}-Q_{\text{off}} \right).
\end{equation}

Finally, we report the cut-off frequency computed for $V_{\rm DS} = V_{\rm DD}$ as

\begin{equation}
f_{T} = \frac{1}{2\pi} \frac{\partial I_{\text{DS}}/\partial V_{\text{GS}}}{\partial Q_{\text{tot}}/\partial V_{\text{GS}}},
\end{equation}

which is a relevant quantity for high frequency analog applications.

From Table \ref{tab1} we can immediately draw that the LH-FET exhibits an almost ideal SS~=~68~-~72~mV/decade and I$_{\text{on}}$/I$_{\text{off}} > 10^4$ for channel lengths $\geq$ 5.5~nm for the HP case, and $>10^6$ for the LP case. As expected, the HP case is optimized for low $\tau$ (reaching 0.22~ps for the 5.5~nm channel length vs. 0.44~ps of the LP case) whereas the LP case is optimized for low PDP (reaching 0.12~fJ/$\mu$m for the 5.5~nm channel length vs. 0.16~fJ/$\mu$m for the HP case).

As the channel length is reduced, we have a monotonous decrease of $\tau$ and PDP and an increase of $f_T$, meaning that both digital and analog performance is improving. However, for channel lengths of 3.3~nm we observe a significant increase of the intra-band tunneling leakage current, which reduces the I$_{\text{on}}$/I$_{\text{off}}$ for the HP process and does not allow to reach the required $I_{\rm off}$ for the LP case.

For comparison purposes, we also consider the performance of the LH-FET for less aggressive gate dielectric thickness (in terms of equivalent silicon oxide thickness). In Figs. \ref{fig:Fig3}c) and \ref{fig:Fig3}d), we report the transfer characteristics for the LH-FET with channel length of 5.5~nm and for $t_{\text{ox}}$~=~0.5, 1.0 and 1.5 nm. The intrinsic delay time and the power delay product for HP case are reported in the inset. As expected, $\tau$ increases and PDP decreases as the oxide thickness is increased.

The second transistor concept we investigate is the planar barristor, shown in Fig. \ref{fig:Fig2}b). We  consider a double gate (DG)
planar device with top and bottom silicon (or equivalent) oxide thickness $t_{\text{ox}}$ = 0.5 nm and a single gate configuration (SG) with bottom oxide thicknesses of  $t_{\text{oxB}}$~=~0.5~nm and $t_{\text{oxB}}$~=~5~nm.

\begin{figure}[h!!!]
\includegraphics[width=8cm]{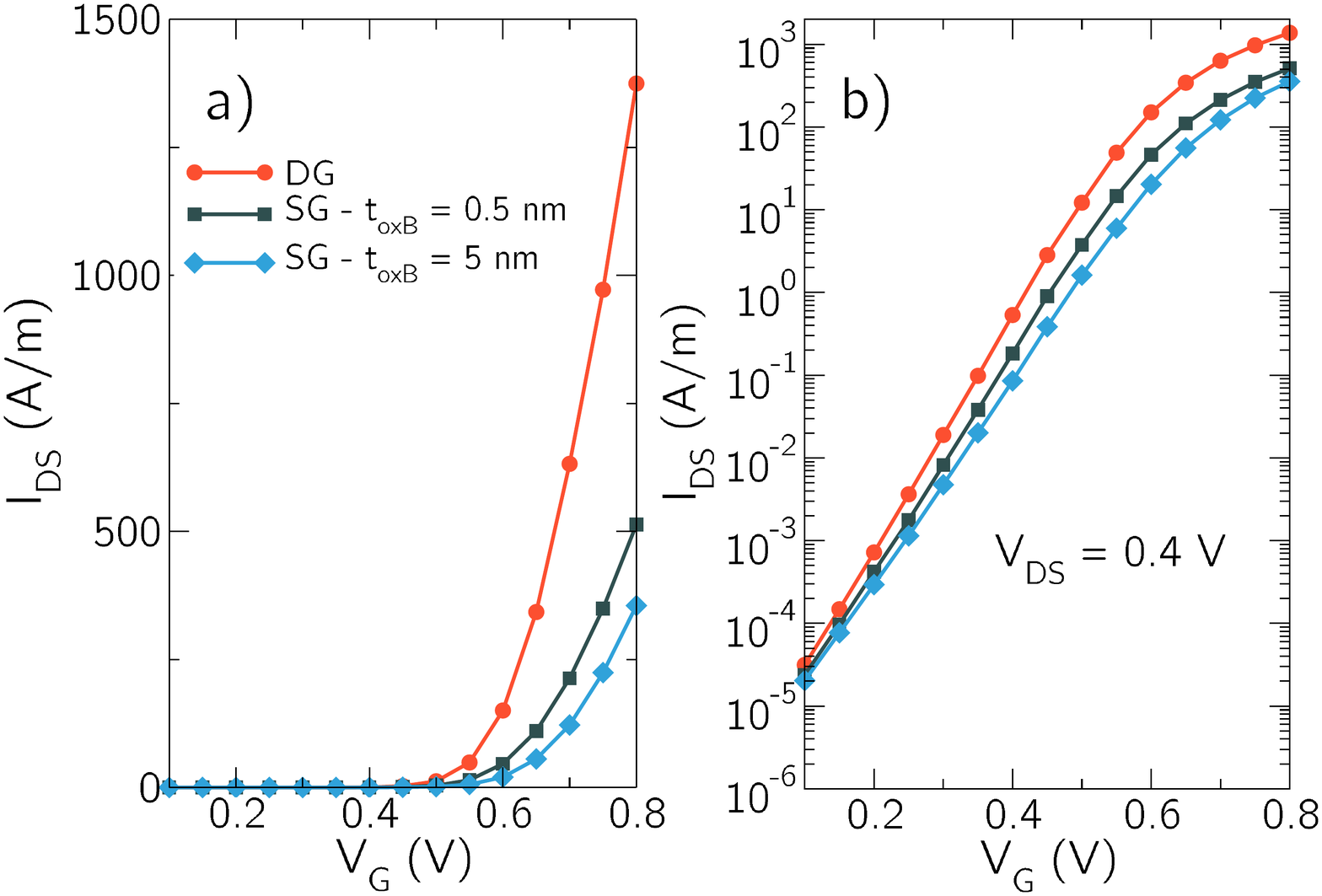}
\caption{Transfer characteristics in a) linear and b) logarithmic scale of the DG planar barristor, and of the SG planar barristors with
$t_{\text{oxB}}$ = 0.5 nm and $t_{\text{oxB}}$~=~5~nm for $V_{\text{DS}}$~=~0.4~V. The devices are illustrated in Fig. \ref{fig:Fig2}b).}
\label{fig:Fig7}
\end{figure}

In Fig. \ref{fig:Fig7}, we show the transfer characteristics for planar barristor with $V_{\text{DS}}$ = 0.4~V. SS is reasonably good (SS~$ < 79$~mV/dec), and the $I_{\text{on}}$/$I_{\text{off}}$ is close to $10^4$ for the high performance case only when the structure is double gated. The same figures of merit computed for the LH-FET, have been calculated for the planar barristor and reported in Table \ref{tab1}. The barristor and LH-FET do not differ much in terms of electrostatic control of the channel/semiconductor region, and therefore both PDP and $\tau$ show similar values for both devices when a double gate geometry is considered. The single-gated barristor shows, however, sightly worse figures of merit, for both HP and LP applications, as compared to the double-gated device, due to the smaller gate-to-channel capacitance.

\begin{figure} [h!!!]
\includegraphics[width=8.5cm]{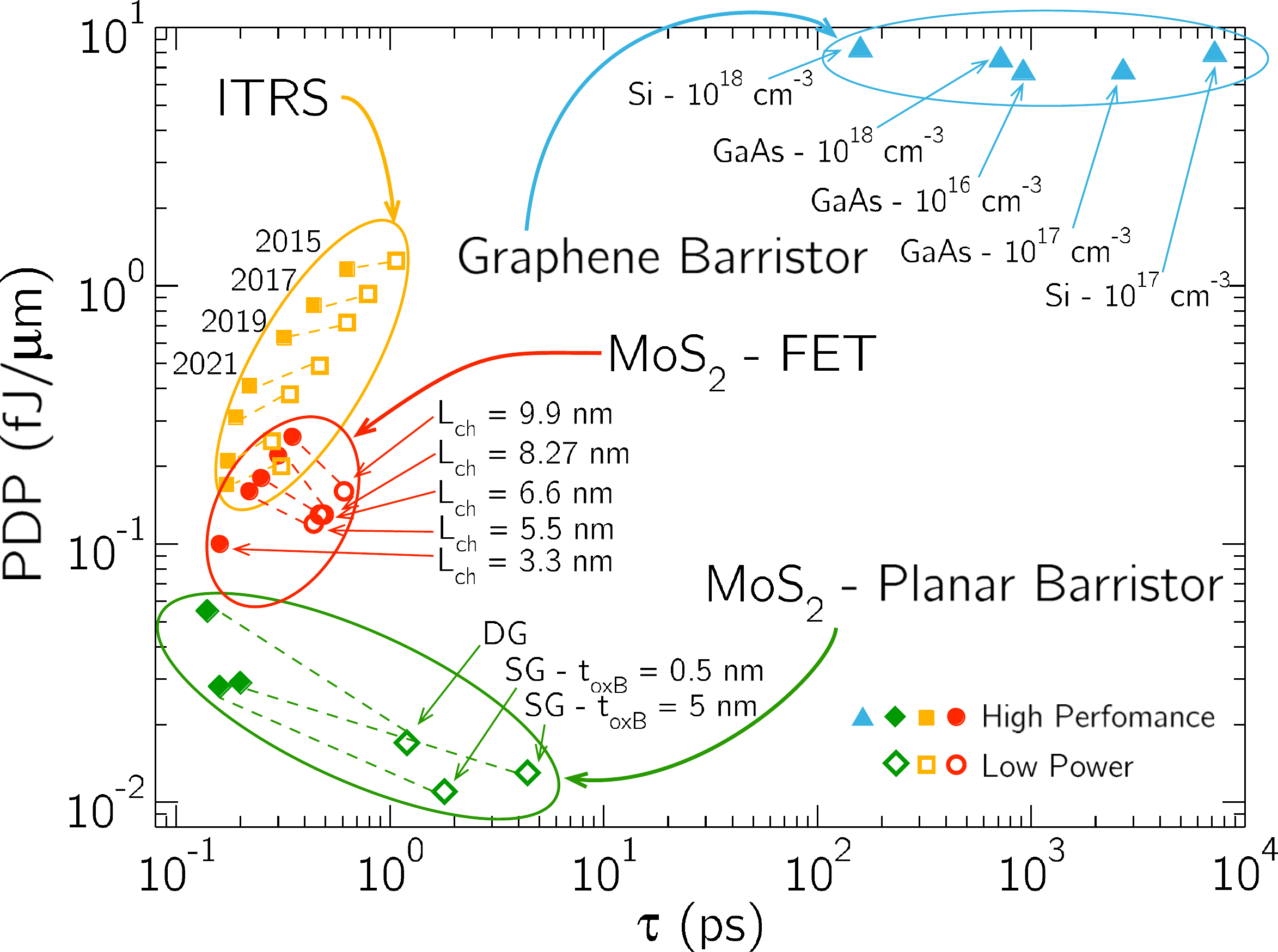}
\caption{PDP and $\tau$ of the MoS$_2$ FETs studied in this work (red circles for the LH-FET and green diamonds for the planar barristor) compared with the requirements of the ITRS 2015 \cite{ITRS} until the end of the roadmap (yellow squares) and the graphene barristors studied in Ref \cite{Logoteta2014} (blue triangles). Filled symbols correspond to figures of merit related to the HP 
process while empty symbols to the LP process. We have assumed for the graphene barristor a width of 1~$\mu$m. In the case of the graphene barristor of Ref \cite{Fiori2012} results for GaAs and Si substrates are shown, for different values of the doping (the concentration is indicated for each device).}
\label{fig:Fig8}
\end{figure}

Finally, in Fig. \ref{fig:Fig8} we compare $\tau$ vs. PDP of the MoS$_2$ FETs studied in this work with predictions from ITRS 2015 \cite{ITRS} for HP and LP CMOS technology and with simulations for the optimized graphene-based barristor studied in Ref. \cite{Fiori2012}. The most desirable region of the plot in Fig. \ref{fig:Fig8} is obviously the bottom left corner.

From the comparison, it appears that the LH-FET and the planar barristor have competitive expected performance with respect to CMOS technology, both for HP and LP applications. The graphene vertical barristor misses the required values for the $\tau$ and PDP figures of merit by several orders of magnitude. As some of us have already discussed \cite{Logoteta2014}, the main problem is that in the vertical barristor the source (the graphene sheet) is between the control gate and the barrier to be modulated, and therefore adds a high parasitic capacitance to the gate, degrading the electrostatics and hence device performance. In the case of the planar barristor, instead, the source is in the plane of the barrier, whereas the control gate is off plane, and therefore the parasitic capacitance is negligible. The proposed results provide an upper limit for the performance of transistors based on heterostructures of MoS$_2$. The actual device performance can be degraded by both intrinsic unavoidable causes, such as phonon scattering, and by technology-dependent factors, such as contact resistance, heterointerface roughness, and  other defects. As of now, the latter factors represent the main performance limitation, even if MoS$_2$ fabrication technology is improving at an impressive pace \cite{Kappera2014,Kappera2014_2,Nourbakhsh2016}, whereas phonon scattering has a limited impact in the case of 5-nm channel length.

\section{Conclusion}

In this work we have presented two different transistor concepts based on planar heterostructures of 1T-MoS$_2$ and 2H-MoS$_2$, that exploit the possiblity of defining with a top-down technique patterns of metallic 1T-MoS$_2$ in a monolayer of semiconducting 2H-MoS$_2$. We have shown with multiscale simulations that the lateral heterostructures FET and the planar barristor exhibit promising performance with respect to the  evolution of CMOS technology predicted by the 2015 edition of the ITRS, both for high performance and for low power applications.

Clearly, our multiscale simulations consider defectless devices with ideal contacts and fully ballistic transport. Therefore, ours is a {\em via negativa} approach: it can only rule out device concepts as good candidates for digital logic if their (ideal) performance is not sufficiently good. 
In our case, it signals that the proposed transistor concepts are extremely promising. Further investigation is required to verify the impact of defects and non-idealities on performance.

\begin{acknowledgments}
Authors gratefully acknowledge the support from the EC 7FP through the GRADE project (agreement number 317839) and the Graphene Flagship (agreement number 696656). Computational resources at nanohub.org are gratefully acknowledged. 
\end{acknowledgments}

\appendix*
\section{Computational Methods}
\label{App}

\begin{figure}[h!!!]
\includegraphics[width=7cm]{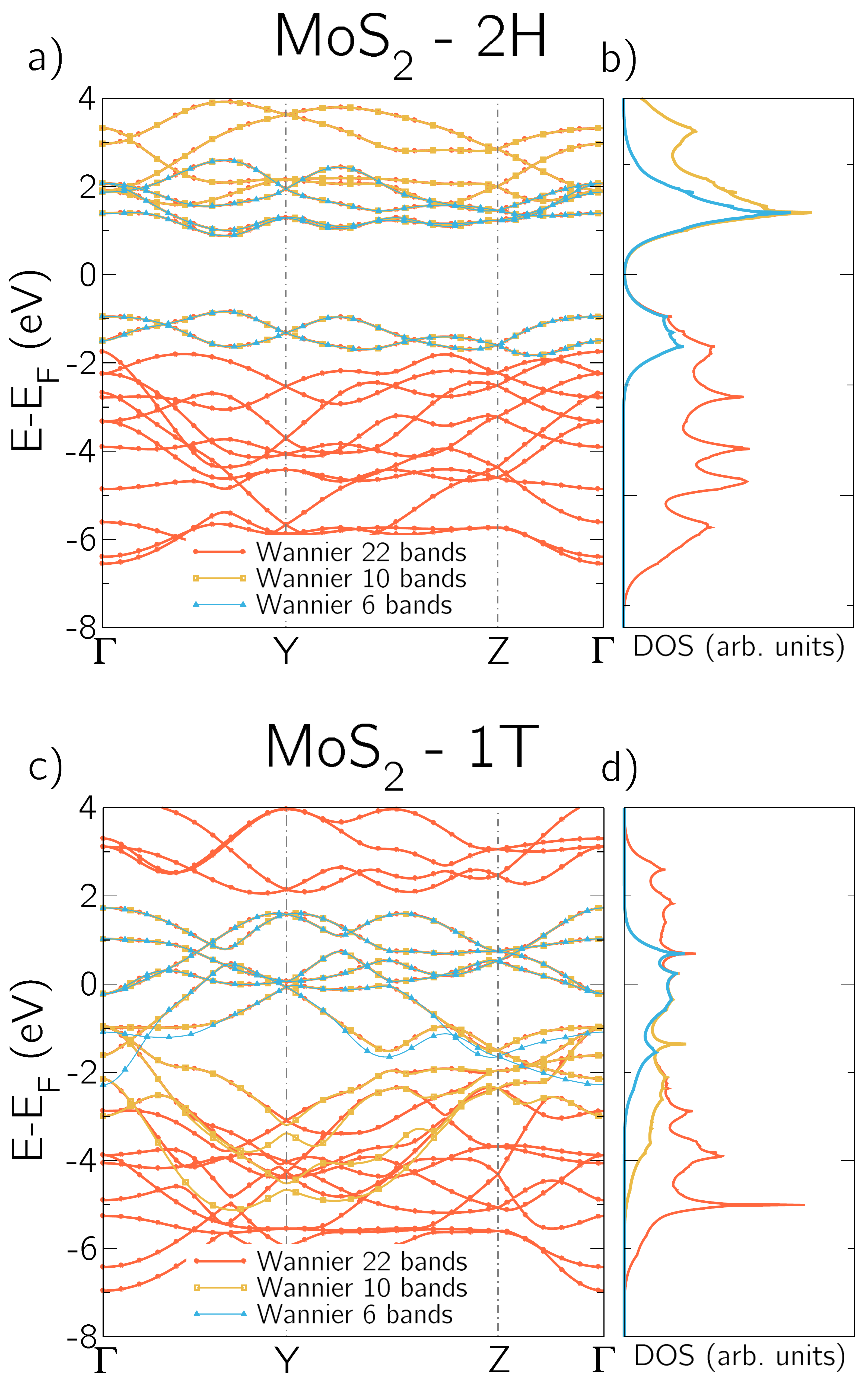}
\caption{Bands computed through MLWF for different numbers of Wannier centers (i.e. number of considered bands) and the corresponding density of states (DOS) for a) and b) MoS$_2$-2H, and c) and d) MoS$_2$-1T.}
\label{fig:Fig1S}
\end{figure}

DFT simulations have been performed with the Quantum Espresso package \cite{QE}. For both materials, MoS$_2$-2H and MoS$_2$-1T, we have considered 8x12x1 Monkhorst-Pack grid which has been evaluated to be sufficiently accurate for the subsequent calculations. We have used ultrasoft pseudopotentials with PBE exchange correlation functional. A 60 Ry wave function cutoff, a 600 Ry charge density cutoff and a vacuum layer of 20 \r{A} have been considered. The calculation of the MLWF has been performed with Wannier90 \cite{Wannier} using the same grid of the DFT calculations. In Figs. \ref{fig:Fig1S}a) and \ref{fig:Fig1S}c), we show the bands for MoS$_2$-2H and MoS$_2$-1T, considering a varying number of Wannier centers (i.e., bands): 6, 10 and 22. It is evident that 6 bands provide the same results as the 22 bands case (but with a smaller computational cost) if we consider an energy range close to the Fermi level, i.e., the energy range taking part in transport. We also report in Figs. \ref{fig:Fig1S}b) and \ref{fig:Fig1S}d) the density of states projected on the different number of Wannier centers. As expected the density of states of 6, 10 and 22 bands provides the very same results close to the Fermi level.

\begin{figure}[h!!!]
\includegraphics[width=9cm]{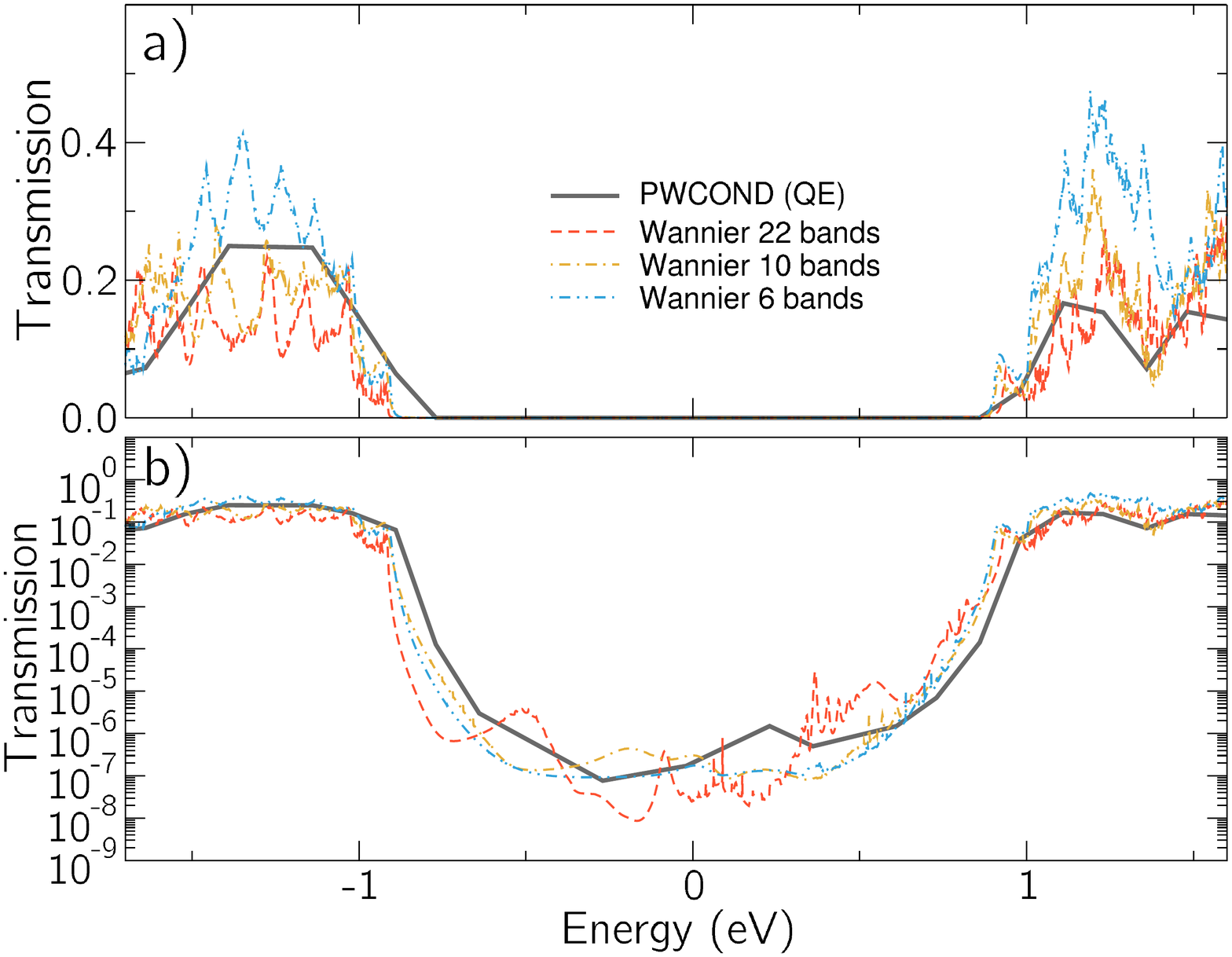}
\caption{Transmission coefficient of the LH-FET device in linear a) and semilogarithmic b) scale, considering a flat potential equal to zero. Comparison between PWCOND results and Wannier Hamiltonian for different number of bands using the Wannier Hamiltonian of 2H-MoS$_2$ as coupling between metallic and semiconducting phases.}
\label{fig:Fig2S}
\end{figure}

To validate the multi-scale approach, we have compared the computed transmission coefficient considering a small number of bands with that obtained from ab-initio calculations (by means of the PWCOND suite of Quantum Espresso \cite{QE}). We have considered a LH-FET with 3.3 nm MoS$_2$-2H channel, as studied in the manuscript. The total size of the system is 7.2 nm, the 1T parts are 1.9 nm and 1.3 nm, the 2H channel is 3.3 nm, while the remaining length corresponds to the interface between the two phases (see Ref. \cite{Cusati2016}). The three MLWF Hamiltonians fit well the PWCOND result, validating the use of the 6-band Hamiltonian (see Fig. \ref{fig:Fig2S}). 

\begin{figure}[h!!!]
\includegraphics[width=8cm]{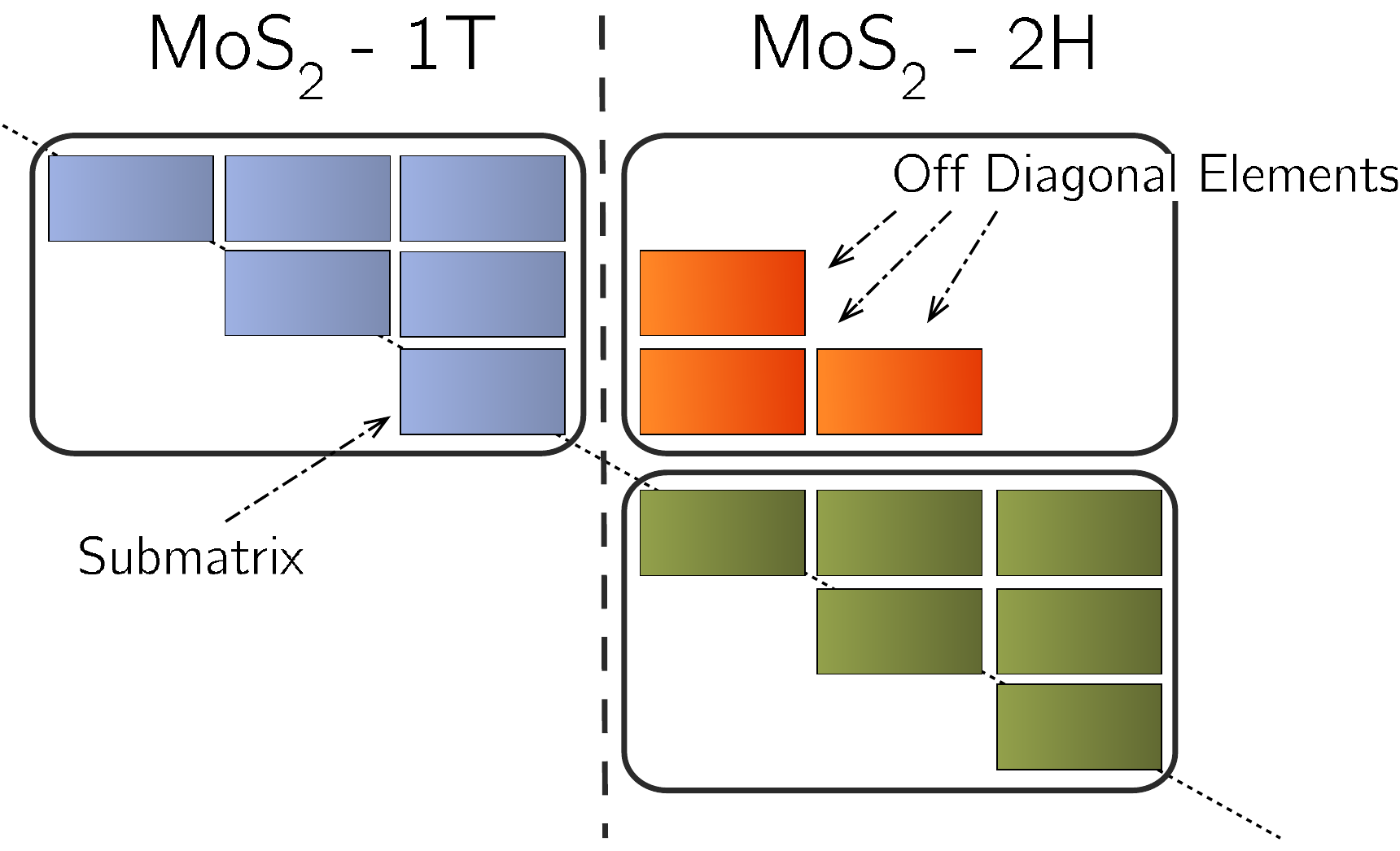}
\caption{Schematic representation of the construction of the MoS$_2$-1T/MoS$_2$-2H interface. The off-diagonal elements, which are colored in red, have been chosen to be equal to the MoS$_2$-2H elements.}
\label{fig:Fig3S}
\end{figure}

In order to build the interface Hamiltonian between the two materials one has to pay attention to the off diagonal elements connecting the two different materials. The heterostructure Hamiltonian for the MoS$_2$-1T/MoS$_2$-2H interface has been determined as follows. In Fig. \ref{fig:Fig3S} we report a schematic representation of the Wannier Hamiltonian along the transport direction. In particular, each submatrix in Fig. \ref{fig:Fig3S}, whose dimension is $N_{wan} \times N_{wan}$ (where $N_{wan}$ is the number of Wannier centres), is connected to the other submatrices along the same row (the number of submatrices along the same row is constrained by the Monkhorst-Pack grid used). Then, one has to choose how to deal with those elements which connect one material to the other (indicated in Fig. \ref{fig:Fig3S} as off diagonal elements with red color). These elements belong to a row where the first element is of the first material (MoS$_2$-1T) and to a column of the second material (MoS$_2$-2H). We decide that these elements to be the ones of the MoS$_2$-2H because of two reasons: firstly, the results of the transmission coefficient reported in Fig. \ref{fig:Fig2S} are in agreement with the PWCOND results, and secondly, among the other possibile choices, the simulations give a better and faster convergence.

\end{document}